\documentclass[letter, twocolumn]{jpsj2} %% for letters
\title{Imprinting the Memory into Paste and its Visualization
 as Crack Patterns
 in Drying Process}

\author{Akio \textsc{Nakahara}
\thanks{E-mail address: nakahara@phys.ge.cst.nihon-u.ac.jp}
 and Yousuke Matsuo}

\inst{Laboratory of Physics, College of Science and Technology \\
Nihon University, Funabashi, Chiba 274-8501, Japan}

\abst{
In the drying process of paste,
 we can imprint into the paste
 the order how it should be broken in the future.
That is,
 if we vibrate the paste before it is dried,
 it remembers the direction of the initial external vibration,
 and
 the morphology of resultant crack patterns is determined
 solely by the memory of the direction.
The morphological phase diagram of crack patterns
 and the rheological measurement of the paste show that
 this memory effect is induced by the plasticity of paste.
}

\kword{memory, crack pattern, yield stress, plasticity, paste}

\begin{document}
\maketitle

\newpage

%\par\noindent
It has been thought that
 crack formations are usually random and stochastic phenomena
 which are easily affected by a change in the cracking process.
For example,
 random crack patterns are formed due to rapid impacts
\cite{Herrmann90, Lawn93, Oddershede93}
 and a transition between regular crack patterns appears
 under controlled experimental situations
\cite{Yuse93, Sasa94}.
In both cases,
 the morphology of crack patterns depends on the cracking condition.
In this letter,
 we report that in the drying process of paste
 we can imprint into the paste
 the order how it should be broken in the future.

% Introduction
% Usual cellular crack pattern & Anisotropic crack pattern
When we mix powder with a lot of water
 to get colloidal suspension,
 pour the water-rich mixture into a container
 and keep it in an air-conditioned room
 of the fixed temperature and humidity,
 crack patterns emerge as the mixture is dried.
In many cases,
 the crack patterns are of isotropic and cellular structures
\cite{Groisman94}.
However,
 when we dry a water-poor mixture such as paste,
 we find that
 we can control the morphology of future crack patterns
 and produce anisotropic crack patterns
 shown in Fig. 1
 by applying the paste an external vibration
 before it is dried
\cite{Nakahara03}.

\begin{figure}
\begin{center}
\includegraphics*[width=8cm]{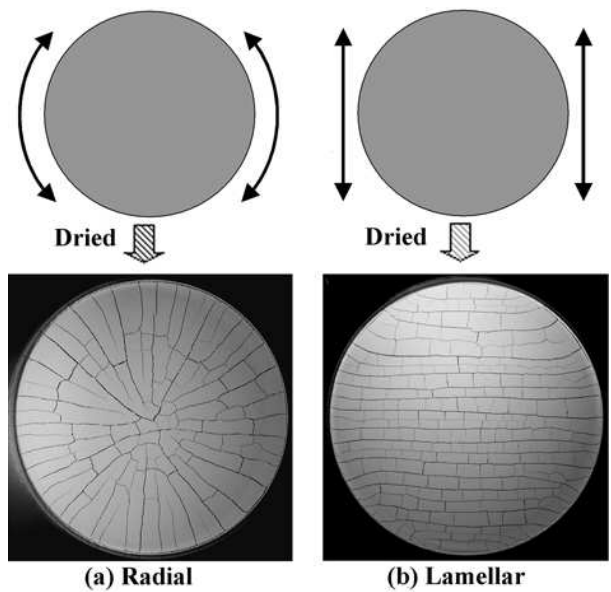}
\end{center}
\caption{Anisotropic crack patterns.
The diameter of each container is 500mm.
 (a) The radial crack pattern
 appears when the container was initially oscillated
 in an angular direction.
 (b) The lamellar crack pattern
 appears when the container was initially vibrated
 in one direction.}
\end{figure}

%Experimental setup
As the powder,
 we use particles of Calcium Carbonate ${\rm CaCO_3}$
 (Kanto Chemical Co., Tokyo).
By performing microscopic observations
 using a Scaninng Electron Microscope (SEM)
 (Hitachi Co., Tokyo),
 we find that
 the shape of dry particles of ${\rm CaCO_3}$ is isotropic and rough,
 and the sizes of particles range
 between $0.5\mu{\rm m}$ and $5\mu{\rm m}$.
The water-poor mixture
 is prepared
 by mixing the 3000g of powder with the 1500g of distilled water.
Since the density of Calcium Carbonate is $2.72{\rm g/cm^3}$,
 the volume fraction of powder in the mixture,
 denoted by $\rho$,
 is given by $\rho=44\%$.
We pour the mixture into an acryl cylindrical container
 of 500mm in diameter.
The thickness of the mixture becomes 13mm.
During the drying process,
 we keep the container at 25${\rm {}^o C}$ and at 30\% in humidity.

% Usual isotropic and cellular crack pattern
Usually,
 an isotropic and cellular crack pattern appears in 3 days
 as the mixture is dried.
The sizes of these fragments are almost the same like cells,
 the shape of each fragment is isotropic,
 and they are distributed randomly in space.
Here,
 let us explain how cracks are formed
 in the drying process.
A thin layer of the mixture of powder and water is contracted
 as the water is evaporated.
If the adhesion between the mixture and the container is weak,
 the mixture slips on the container during the drying process,
 so that the mixture can shrink homogeneously and no cracks are formed.
We perform drying experiments using a tefron-coated container
 and confirm that,
 when the adhesion is weak,
 the mixture shrinks homogeneously with no cracks.
On the other hand,
 if the adhesion between the mixture and the container is strong,
 the mixture sticks to the bottom of the container,
 so the contraction of the mixture in the horizontal direction
 is prohibited
 unless the mixture is split into smaller pieces by the formation of cracks.
The subsequent formation of new cracks
 continues
 until the sizes of fragments become
 about the thickness of the mixture.
In fact,
 the characteristic sizes of these fragments are
 reported to be proportional to the thickness of the mixture
\cite{Groisman94, Allain95, Komatsu97, Kitsune99}.

%\begin{figure}[tb]
%%\begin{center}
%%\includegraphics{}
%%\end{center}
%\caption{Basically, figures and tables must be located near the place where they appear for the first time in the text.}
%\label{f1}
%\end{figure}

% First experiment on lamellar crack pattern
To our surprise,
 when we oscillate the container in an angular direction
 at the frequency of 60rpm and at the amplitude of 0.1rad
 for 10sec just after we pour the mixture into the container,
 a radial crack pattern emerges,
 as is shown in Fig. 1(a),
 with the direction of these radial cracks
 all perpendicular to the angular direction of the cylindrical container.
One might conjecture that
 the radial crack pattern is formed
 due to the shape of the cylindrical container.
However,
 when we vibrate the cylindrical container in one direction for 10sec
 at the frequency of 60rpm and at the amplitude of 15mm
 as is shown in Fig. 1(b),
 a lamellar crack pattern emerges as the mixture is dried,
 and the direction of lamellar cracks is
 perpendicular to the direction of the initial external vibration
 and does not depend on the shape of the container.
That is,
 the water-poor mixture
 remembers the direction of the initial external vibration
 and its memory is visualized
 as the morphology of anisotropic crack patterns.

Do these memory effects apply
 only to the mixture of water and powder of Calcium Carbonate?
Or, do these memory effects also
 apply to mixtures of water and other powders?
In the experiments using clay particles
 such as Kaolin,
 we find that, when we vibrate and then dry clay,
 we obtain similar anisotropic crack patterns
 which reflect the direction of the initial external vibration.
%Even a paste of fine carbon particles
% is found to show the memory effect of the external vibration.

Why do these water-poor mixtures remember
 the direction of the initial external vibration?
First,
 we consider that
 the volume fraction of powder in the mixture, $\rho$,
 plays an important role in the memory effect.
When the volume fraction of particles is small,
 particles are distributed randomly and sparsely,
 so that the mixture is regarded as a Newtonian viscous fluid.
On the other hand,
 when the volume fraction of particles exceeds a threshold value,
 a structure or a network of particles is formed
 based on the van der Waals attractive interaction,
 and the rheological feature of the mixture changes drastically.
This very concentrated suspension of solid colloidal particles
 is regarded as a paste
\cite{Blanc95}.

Next,
 we notice that,
 to imprint the memory into the mixture,
 we should apply an initial external vibration
 with an appropriate strength.
Here,
 the strength of the vibration is given by $4 \pi^2 r f^2$,
 where $r$ is the amplitude and $f$ is the frequency
 of the vibration.
Thus,
 in the following experiments,
 we systematically change the values of two main parameters,
 i.e.,
 the volume fraction $\rho$ of powder in the mixture
 and the strength $4 \pi^2 r f^2$ of the initial external vibration,
 to investigate the mechanism of imprinting the memory into paste.

The shapes of acryl containers are square boxes
 with 200mm in length each side.
We fix the mass of powder in the mixture as 360g in each container,
 so that we can equalize
 the final thickness of mixtures with different volume fractions
 when they dry up
 and thus equalize characteristic sizes of final crack patterns.
For example,
 the final thickness of the mixture
 when we use the 360g of the powder is $ 6 \sim 7 {\rm mm}$.
As long as we perform experiments using the mixture with its thickness
 larger than the thickness of the boundary layer, 5mm,
 we find that we obtain same experimental results.
To change the value of the volume fraction $\rho$,
 we change the amount of water to mix with.
After we pour the mixture into the container
 which is set on a Shaker FNX-220 (TGK Co., Tokyo)
 or a Triple Shaker NR-80 (Taitec Co., Tokyo),
 the container is vibrated horizontally in one direction for 60sec,
 then we stop the vibration and wait for cracks to appear.
To change the value of the strength of the vibration,
 we change the value of the frequency $f$ from 20 to 60rpm
 and also the value of the amplitude $r$ as 10, 15, and 20mm.

We show a morphological phase diagram of the crack patterns,
 in Fig. 2,
 as a function of the volume fraction $\rho$ of powder in the mixture
 and the strength $4 \pi^2 r f^2$ of the initial external vibration.
In Fig. 2,
 the amplitude $r$ of the vibration is set to be 15mm
 and only the value of the frequency $f$ is varied,
 but, we also get the same results
 as we set the value of the amplitude $r$ as 10mm and 20mm.
When the value of $\rho$ is smaller than $\rho=25\%$
 which is denoted by the vertical dotted line,
 only isotropic cellular crack patterns appear.
When the value of $\rho$ is larger than $\rho=54\%$
 which is denoted by the vertical dashed-and-dotted line,
 we cannot make a morphological phase diagram,
 because we cannot mix powder with water homogeneously
 due to the lack of sufficient water.
The region between these two lines
 is divided,
 by the solid and the dashed curves,
 into three regions, A, B, and C.
Only cellular crack patterns appear
 in regions A and C,
 where, in the region A,
 the value of the strength $4 \pi^2 r f^2$
 of the initial external vibration is
 smaller than that of the solid curve,
 and, in the region C,
 the value of $4 \pi^2 r f^2$ is larger than that of the dashed curve.
The method to draw these two guide curves,
 i.e., the solid curve and the dashed curve,
 will be explained in the forthcoming paragraphs.
Only in the region B between the solid and the dashed curves,
 we find anisotropic lamellar crack patterns
 with the memory of the initial external vibration.

\begin{figure}
\begin{center}
\includegraphics*[width=7.62cm]{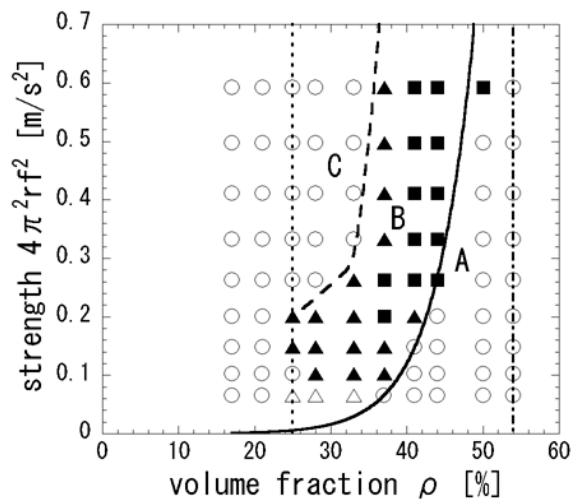}
\end{center}
\caption{Morphological phase diagram of crack patterns
 as a function of the volume fraction $\rho$ of powder in the mixture
 and the strength $4 \pi^2 r f^2$ of the initial external vibration.
Open circles denote isotropic cellular crack patterns,
 solid squares denote lamellar crack patterns,
 and solid triangles denote
 combinations of isotropic cellular crack patterns
 and lamellar crack patterns.
As for open triangles,
 we get isotropic cellular crack patterns,
 but, when we use larger square boxes
 with 400 mm in length each side
 and the 1440 g of powder,
 we get combinations of
 isotropic cellular crack patterns and lamellar crack patterns.
This is due to the finite size effect,
 because the side walls prevent the movement of the mixture
 especially when the strength of the external vibration is weak,
 and thus prevent the memorizing process.
The dotted and the dashed-and-dotted lines correspond to
 $\rho=25$ and $54\%$, respectively.
The region between these two guide lines are divided,
 by the solid and the dashed curves,
 into three regions, A, B, and C.
Only in the region B between the solid and the dashed curves,
 lamellar crack patterns can appear.
}
\end{figure}

How can the mixture remember the initial vibration
 when it is in the region B?
At first,
 we suspected that
 the initial vibration caused
 wrinkles on the surface of the mixture,
 which became triggers
 to form anisotropic crack patterns.
However,
 when we observe
 the surface of the mixture after the initial vibration,
 the surface is flat and smooth
 and we cannot see any wrinkles nor any traces of the initial vibration.
To avoid any effect of the as-formed surface condition,
 we cut and remove the upper one third of the pouring
 when the mixture becomes a bit harder
 12 hours after we poured the mixture into the container.
Resultant crack patterns are, however,
 not affected by slicing off the top.
That is,
 the memory of the initial vibration
 must be kept somewhere inside the mixture.

Next,
 we have investigated the rheology of the mixture
 by using Dynamic Stress Rheometer (DSR)
 (Rheometrics Co., Riscataway, NJ).
The mixture of 0.4ml in volume is spread on a plane plate of DSR,
 and is sandwiched between the plate
 and a cone plate of 25mm in diameter and 0.1rad in angle.
Just after we sandwich the mixture between these two plates,
 we perform a step stress creep test.
For each step,
 we keep the value of the shearing stress  $\sigma$ for 60sec,
 and measure the asymptotic value of the shear rate ${\dot{\gamma}}$.
Graphing the shearing stress $\sigma$
 as a function of the shear rate ${\dot{\gamma}}$,
 we see a linear relation between $\sigma$ and ${\dot{\gamma}}$,
 and the intercept at a vertical axis
 corresponds to its yield stress $\sigma_Y$.
In Fig. 3,
 we present the result of the yield stress $\sigma_Y$ of the mixture
 as a function of the volume fraction $\rho$.
We find that,
 when the value of $\rho$ is less than 25\%,
 the mixture is regarded as a Newtonian viscous fluid
 with a vanishing yield stress $\sigma_Y$.
When the value of $\rho$ becomes greater that 25\%,
 the value of $\sigma_Y$ increases drastically
 as the value of $\rho$ increases,
 and the mixture is described by a visco-plastic fluid
 with a non-zero yield stress $\sigma_Y$.
This drastic increase of the yield stress of the mixture
 of water and powder of Calcium Carbonate
 shows the concentration dependence
 similar to that of clay pastes
\cite{Blanc95}.
When the value of $\rho$ exceeds 54\%,
 the mixture becomes semisolid,
 and
 we cannot measure the value of the yield stress any more.

Comparing these results
 with the morphological phase diagram shown in Fig. 2,
 we find that,
 only when the mixture is water-poor and visco-plastic like paste
 with a non-zero yield stress,
 the mixture can remember the direction of the initial external vibration
 and the anisotropic crack patterns appear.
Recently,
 the memory effects are widely observed in dissipative microstructures,
 such as sand piles
\cite{Vanel99},
 microgel pastes
\cite{Cloitre00}
 and rubbers
\cite{Miyamoto02}.
Theoretical approaches based on the elastic-plastic deformation of paste
 with a non-zero yield stress
 are under investigation
 to explain the memory effect of an external mechanical force in paste
 and the formation of resultant anisotropic crack patterns
 in the drying process of paste
\cite{Ooshida04, Otsuki04}.

\begin{figure}
\begin{center}
\includegraphics*[width=7.62cm]{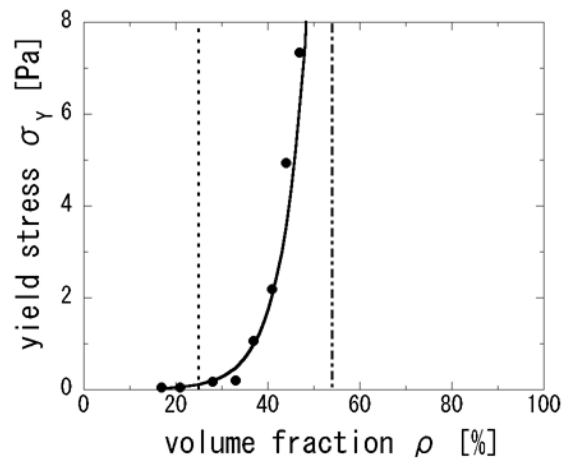}
\end{center}
\caption{Yield stress $\sigma_Y$ as a function
 of the volume fraction $\rho$ of powder in the mixture.
The dotted and the dashed-and-dotted guide lines correspond to
 $\rho=25$ and $54\%$, respectively.}
\end{figure}

There remain some questions.
In Fig. 2,
 we see that,
 even when the mixture is visco-plastic like paste,
 lamellar crack patterns do not appear
 in regions A and C.
We conjecture that,
 to imprint the direction of the initial external vibration into the mixture,
 we should vibrate the mixture
 with an appropriate strength.
The solid curve in Fig. 2 which divides regions A and B is drawn
 by equalizing the value of the shearing stress $\sigma$
 induced by the external vibration
 to that of the yield stress $\sigma_Y$ shown in Fig. 3.
The shearing stress $\sigma$ is expressed
 as an external force acting horizontally
 at the bottom of the container,
 normalized by its area $S$,
 and is expressed as
 $ \sigma = ( m h S ) ( 4 \pi^2 r f^2 ) / S $,
 where $m$ is a density and $h$ is a thickness of the mixture.
In the region A
 where
 the value of the shearing stress $\sigma$
 is smaller than
 that of the yield stress $\sigma_Y$,
 the mixture is not vibrated effectively
 by the initial external vibration.
In fact,
 we observe no movement of the mixture at the initial vibration
 when compared with the movement of the container.
Thus,
 only isotropic cellular crack patterns appear in the region A.
The dashed curve in Fig. 2 which divides regions B and C
 is drawn by checking the fluidity of the mixture at the initial vibration.
In the region C
 where the strength of the initial external vibration is large,
 there appear some surface waves or turbulent flows
 during the initial vibration.
We consider that,
 once a macroscopic flow emerges in the mixture,
 the flow blows up the microscopic memory inside the mixture,
 thus only isotropic cellular crack patterns appear in the region C.
Only in the region B where
 the strength of the vibration is appropriate
 and the shear movement of the visco-plastic mixture is observed,
 the paste remembers the direction of the initial external vibration
 and, as a result, anisotropic lamellar crack patterns emerge.

We wonder how long the mixture can remember the initial external vibration.
We set the container in a box,
 pour the visco-plastic mixture of $\rho = 44 \%$ into the container,
 close the box,
 vibrate the whole system at $f=60{\rm rpm}$ and $r=15{\rm mm}$ for 60sec
 and let it sit for one month.
Since the humidity in the box soon saturates,
 the drying process stops so that no cracks are formed.
After we open the box one month later,
 the drying process proceeds,
 and there appear lamellar cracks,
 the direction of which is perpendicular
 to the vibration that we gave one month ago.

Finally,
 we would like to discuss
 how memories of the initial external vibration remain in pastes.
When we vibrate the very concentrated suspension
 of solid colloidal particles,
 we consider that
 the inelastic collisions between colloidal particles are induced
 by the vibration
 and a density fluctuation of colloidal particles emerges.
It is similar to the formation of density waves or jam structures
 which appear in dense granular flows
\cite{Nakahara97}.
To observe the density fluctuation,
 we perform microscopic observations using microscope and SEM,
 but it is difficult to observe the density fluctuation directly.
This is because colloidal particles are packed densely and randomly in paste.
Even if the amplitude of this density fluctuation is small,
 however,
 it influences the strength and the number of bonds
 which bind neighboring particles
 in the network structure of paste.
Once the paste is dried,
 the weaker bonds at less dense regions
 must be broken earlier.
We consider that the initial external vibration causes
 spatial fluctuation
 of the strength of binding bonds
 along the direction of the vibration,
 and even if the spatial fluctuation is faint to observe,
 it can be visualized easily and clearly
 as an anisotropic crack pattern which appears in the drying process.

We think our experimental results can be applied to industrial sciences,
 because,
 if we can control the direction in which cracks will be formed in future,
 we can make plans to avoid accidental serious damages.
Moreover,
 as no one knows how cracks will be formed
 except those who had imprinted the order,
 this method can be used
 to hide information physically.

To summarize,
 we experimentally found that
 we can imprint the direction of the external vibration into paste,
 and the memory in the paste is visualized
 as the morphology of anisotropic crack patterns
 which appear in the drying process.

We would like to acknowledge
 H. Uematsu, M. Otsuki, S. Sasa, T. S. Komatsu, and Y. Nakahara
 for valuable discussions.
We thank
 M. Sugimoto, T. Taniguchi and K. Koyama
 for support with rheological measurements
 at Venture Business Laboratory of Yamagata University.
We also thank
 Y. Aoyagi, A. Taguchi, K. Nakagawa and A. Itoh
 for support with microscopic observations using SEM
 at Advanced Materials Science Center of Nihon University.

\end{document}